\pdfoutput=1

\documentclass[11pt]{article}

\usepackage[final]{acl}

\usepackage{times}
\usepackage{latexsym}

\usepackage[T1]{fontenc}

\usepackage[utf8]{inputenc}

\usepackage{microtype}
\usepackage{multirow}
\usepackage{acronym}
\usepackage[pdftex]{graphicx}
\usepackage{hhline}

\acrodef{mAP}{mean average precision}
\acrodef{MR}{mean rank}
\acrodef{P@1}{precision at one}
\acrodef{SHS}{Second Hand Songs}
\acrodef{CSD}{Cover Song Detection}
\acrodef{WER}{word error rate}

\title{Innovations in Cover Song Detection: A Lyrics-Based Approach}

\author{
 \textbf{Maximilian Balluff\textsuperscript{1}},
 \textbf{Peter Mandl\textsuperscript{1}},
 \textbf{Christian Wolff\textsuperscript{2}},
\\
\\
 \textsuperscript{1}Hochschule Muenchen University of Applied Science,
 \textsuperscript{2}University of Regensburg,
\\
 \small{
   \textbf{Correspondence:} \href{mailto:maximilian.balluff@hm.edu}{maximilian.balluff@hm.edu}
 }
}

\begin{document}
\maketitle
\begin{abstract}
  Cover songs are alternate versions of a song by a different artist. Long being a vital part of the music industry, cover songs significantly influence music culture and are commonly heard in public venues. The rise of online music platforms has further increased their prevalence, often as background music or video soundtracks. While current automatic identification methods serve adequately for original songs, they are less effective with cover songs, primarily because cover versions often significantly deviate from the original compositions. In this paper, we propose a novel method for cover song detection that utilizes the lyrics of a song. We introduce a new dataset for cover songs and their corresponding originals. The dataset contains 5078 cover songs and 2828 original songs. In contrast to other cover song datasets, it contains the annotated lyrics for the original song and the cover song. We evaluate our method on this dataset and compare it with multiple baseline approaches. Our results show that our method outperforms the baseline approaches.
\end{abstract}

\section{Introduction}
Cover versions, or cover songs, represent reinterpretations of original compositions by different artists, holding a significant place in the music industry and influencing musical culture. These renditions are frequently featured in public venues such as bars, clubs, festivals, and restaurants. The rise of online music platforms has further amplified their ubiquity, often serving as background music or soundtracks for videos \cite{vervielfaeltigungsrechteDiskothekenMonitoring}.

Identifying and monetizing cover songs poses a significant challenge for rights holders, leading to an increasing demand for efficient methods to detect cover songs on a large scale. Composers and performers, often members of collecting societies, rely on accurate song identification for royalty distribution. While existing automatic identification methods effectively handle original songs, detecting cover songs remains challenging due to the substantial variations introduced by different artists in their renditions \cite{ShazamMusikEntdecken}.

\section{Related Work}
\ac{CSD} plays a pivotal role in Music Information Retrieval, especially in recent years. Existing approaches are predominantly centered around audio analysis. For instance, Tsai et al. \cite{tsaiQueryByExampleTechniqueRetrieving2005} focused on tempo, transposition, and accompaniment to compare original songs with their covers. Serrà introduced a method using audio features to compute a similarity matrix between original and cover songs \cite{serrajuliaMusicSimilarityBased2007}. In another study \cite{duBytecoverCoverSong2021}, advanced song embeddings were generated using a convolutional neural network and utilized for similarity computation. These methods commonly extract audio features, ranging from high-level attributes like tempo and beats per minute \cite{tsaiQueryByExampleTechniqueRetrieving2005} to low-level features like Harmonic Pitch Class Profile \cite{serrajuliaMusicSimilarityBased2007} and Constant-Q-Transform \cite{duBytecoverCoverSong2021}. Some approaches combine both, incorporating features like the mean and standard deviation of Mel-Frequency Cepstral Coefficient \cite{yuUsingExactLocality2008}.

In contrast, only a few methods leverage song lyrics for \ac{CSD}. Correya et al. employed a unique approach using the lyrics and title of songs through a bag-of-words algorithm \cite{correyaLargeScaleCoverSong2018}. Similarly, Vaglio et al. proposed a method combining lyrics and audio for cover song detection \cite{vaglioWORDSREMAINSAME2021}. They processed the audio in two branches—one for lyrics and the other for audio. After detecting singing voice, lyrics were transcribed, and string matching identified corresponding cover versions. The results from both branches were then combined for the final output.

\section{Idea and Dataset}

Our central proposition asserts that cover song lyrics exhibit notable similarities with their original counterparts. Given the pivotal role of lyrics in music, especially in genres like Pop, Hip Hop/Rap, Rock/Hardrock/Heavy Metal, and Kids Music, where over 70\% of sales are concentrated \cite{bundesverbandmusikindustriee.v.MusikindustrieZahlenIm2021}, investigating this hypothesis becomes imperative.

To substantiate our claim, we introduce a novel dataset comprising 5078 cover songs and 2828 original songs. Collected from genius.com between June and September 2023, this dataset stands out by providing annotated lyrics for both the original and cover songs. Language analysis, conducted using a transformer for language detection \cite{conneauUnsupervisedCrosslingualRepresentation2020}, reveals that approximately 85\% of the original songs are in English, followed by Spanish (4.7\%) and Portuguese (2.3\%).

\begin{figure}
  \centering
  \includegraphics[width=0.45\textwidth]{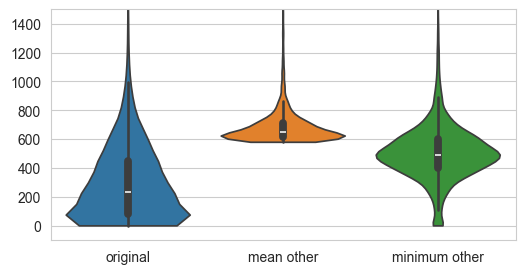}
  \includegraphics[width=0.45\textwidth]{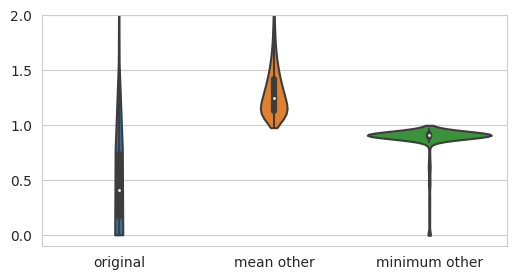}
  \caption{Levenshtein distance between (upper) and \ac{WER} (lower) between cover and original songs, and between cover and other songs}
  \label{fig:lev_distance}
\end{figure}

To assess lyrical similarity, we employ Levenshtein distance \cite{levenshteinBinaryCodesCapable1966} and \ac{WER} \cite{KLAKOW200219}. The mean Levenshtein distance between a cover song and its original 302.62 is notably lower than the mean (689.25) and minimum (504.76) distances to other songs; for more details, compare figure \ref{fig:lev_distance}. Surprisingly, 23.6\% of non-matching songs exhibit a lower distance than the actual original, with 301 instances where the detected language of the cover and original differs. Further analysis reveals instances of low distances in non-matching songs, often attributed to poor annotations, including incomplete cover lyrics or variations in annotation styles (e.g., filler words like "Uh, uh, uh..." or paragraph descriptions like [Verse], [Refrain]). The average \ac{WER} distance between a cover song and its original, denoted as 0.54, is significantly less than the average (1.33) and minimum (0.87) distances observed with other songs. These findings underscore the challenges posed by these differences for classical distance metrics, necessitating thorough preprocessing for accurate analysis.

The dataset will be published in the future. Currently, we are working on increasing the size of the dataset and improving the quality. If possible we will merge our dataset with other datasets like \ac{SHS} \cite{SecondHandSongsDatasetMillion}.

\subsection{Baseline}
As we are the first to utilize our new dataset, our primary objective is to establish multiple baseline algorithms for comparison purposes. We utilize the string similarity distances, namely Levenshtein distance ($L_D$) and \ac{WER} ($WER$), along with a Bag-of-Words algorithm combined with Cosine distance ($BOW$), to calculate the similarity between lyrics. We opted for this set of distances due to their relatively low complexity, and we benefit from existing implementations for Levenshtein and \ac{WER} \cite{levenshtein_github, pytorch_torcheval}.

\subsection{Proposed method}
Since the initial introduction of transformer neural networks \cite{NIPS2017_3f5ee243}, these networks have become the standard for text preprocessing. We utilize a transformer as a feature extractor for the lyrics, which generates an embedding. Lyrics with similar content should yield similar embeddings. The embeddings for each original song are stored in a database. For each cover song, we compute the similarity between the cover song's embedding and the embeddings of the original songs. We order the songs by their similarity, the original song with the highest similarity score is predicted to be the original song for the cover version.

\subsubsection{Model}
\begin{figure}
  \centering
  \includegraphics[width=0.45\textwidth]{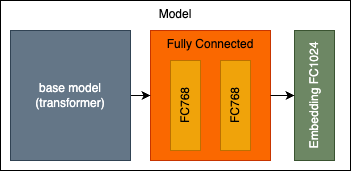}
  \caption{Model architecture}
  \label{fig:model}
\end{figure}

In our approach, we use a pre-trained model for sequence similarity \cite{SymantoSnxlmrobertabasesnlimnlianlixnliHugging}, which is based on XLM-RoBERTa \cite{conneauUnsupervisedCrosslingualRepresentation2020} and is pre-trained on the SNLI \cite{emnlp2015}, MNLI \cite{MultiNLI}, ANLI \cite{AdversarialNLI2023} and XNLI \cite{XNLICrossLingualNLI2023} datasets. On top of the base model, we add two fully connected and an embedding layer. The embedding layer serves the purpose of transforming the sequence of features into a fixed-length floating-point embedding vector, providing a condensed representation of the lyrics that enables subsequent similarity calculations. The model architecture is shown in figure \ref{fig:model}. We did not further optimize the model architecture. The dimensions of the layers are based on our experience in previous text matching projects.

Various methods can be employed to train embeddings, and a common approach is to use Siamese neural networks \cite{chiccoSiameseNeuralNetworks2021}. These networks consist of two identical neural networks that share the same weights. The inputs for these networks are the two samples that need to be compared, with the objective of minimizing the distance between similar inputs and maximizing the distance between dissimilar ones.

In our approach, we adopt the triplet loss method, originally introduced for face recognition \cite{chechikLargeScaleOnline2009}. The goal is to minimize the distance between the anchor and the positive sample while maximizing the distance between the anchor and the negative sample. Achieving this involves sampling triplets, which consist of an anchor, a positive sample, and a negative sample \cite{chechikLargeScaleOnline2009}. To address the challenge of the relatively high cost of finding good triplets, we employ the online mining strategy \cite{schroffFaceNetUnifiedEmbedding2015}. Instead of mining triplets before passing them to the model, the triplets are mined within the training batch. This is accomplished by passing a batch of samples to the model, calculating the embeddings and distances for all embeddings in the batch, and then searching for each sample used as an anchor for other positive and negative samples within the batch.

\subsubsection{Evaluation}
The evaluation of our approach relies on three widely-applied metrics in the field of Cover Song Detection (\ac{CSD}): \ac{mAP}, \ac{MR}, and \ac{P@1}. These metrics have been extensively used in the evaluation of various \ac{CSD} approaches \cite{duBytecoverCoverSong2021, duBytecover2DimensionalityReduction2022, chiccoSiameseNeuralNetworks2021}. \ac{mAP} calculates the average precision of a ranked list of documents, where precision is defined as the ratio of relevant documents among the retrieved documents \cite{manningchristopherIntroductionInformationRetrieval2008}. \ac{MR} provides the position of the first relevant document in the ranked list of documents, with lower values indicating better performance (ideally one) \cite{osmalskyjCombiningApproachCover2017}. \ac{P@1} represents the precision at the first rank of the ranked list of documents \cite{xuKeyInvariantConvolutionalNeural2018}.

Given the limited size of our dataset, we chose to implement k-fold cross-validation with five folds, as recommended by \cite{DataMiningPractical2011}. In order to maintain consistency and comparability across all experiments, the same folds were utilized. Each fold encompassed several cover lyrics and their corresponding originals. The fold was further divided into training (64\%), validation (16\%), and testing (20\%) sets. The validation set is necessary to prevent our model from overfitting the training data, using EarlyStopping and ReduceLearningRate callbacks that monitore the validation loss.

It is crucial to note that no original version is duplicated among these sets, ensuring that each set contains a unique set of original songs. This resulted in varying sizes for each fold, due to the uneven distribution of the number of covers per original. Table \ref{tab:folds} provides an overview of the number of unique lyrics (original and cover) for each fold.

To evaluate the model's performance, we utilized the test set. For each lyric, whether it be a cover or an original, we generated embeddings. The embeddings from the original lyrics formed our database. Subsequently, for each cover embedding, we calculated the Manhattan distance \cite{royerSimultaneOptimierungProduktionsstandorten2001} to all original embeddings, sorting them based on their distances. Our metrics were then computed using this ranking. Model training was conducted on an Nvidia T4 GPU, utilizing a n1-standard-8 instance on Google Cloud. The training process spanned approximately 22 hours for all five folds, averaging around 19 epochs. Notably, our model boasts 278 million parameters, of which 831 thousand are trainable.

\begin{table}[]
  \centering
  \begin{tabular}{l|l|l|l|}
    \cline{2-4}
                                          & \textbf{Training} & \textbf{Validation} & \textbf{Test} \\ \cline{1-4}
    \multicolumn{1}{|l|}{\textbf{Fold 1}} & 5058              & 1313                & 1535          \\ \hline
    \multicolumn{1}{|l|}{\textbf{Fold 2}} & 4975              & 1397                & 1534          \\ \hline
    \multicolumn{1}{|l|}{\textbf{Fold 3}} & 5069              & 1224                & 1613          \\ \hline
    \multicolumn{1}{|l|}{\textbf{Fold 4}} & 4953              & 1334                & 1619          \\ \hline
    \multicolumn{1}{|l|}{\textbf{Fold 5}} & 5026              & 1275                & 1605          \\ \hline
  \end{tabular}
  \caption{Unique lyrics (original and cover) for each fold}
  \label{tab:folds}
\end{table}

Our baseline approaches do not require training, and therefore, we only use the test set for evaluation. We compute the distance between the cover song and all original songs. The original song with the smallest distance is then predicted as the original for the cover song.

\subsubsection{Results}
The results of our experiments are summarized in Table \ref{tab:results}. The Bag-of-Words ($BOW$) approach emerges as the best baseline, yielding a \ac{mAP} of 85.74\%, a \ac{MR} of 46.29, and Precision@1 of 83.65\%. The triplet model outperforms the baseline approaches in all metrics except precision, achieving a \ac{mAP} of 87.17\%, a \ac{MR} of 18.51, and Precision@1 of 83.57\%. Notably, all methods yield approximately the same \ac{mAP}, while the \ac{MR} varies significantly. Remarkably, the triplet model, with its substantially lower rank, outperforms the $BOW$ model by more than a factor of two.

Although we did not initially plan to evaluate the computation time of the approaches, it is worth noting that \ac{WER} distance is significantly slower. It took over 8 hours to compute the distance for all splits, whereas for all other approaches, it was a matter of minutes.

\begin{table}[]
  \begin{tabular}{l|r|r|r|}
    \cline{2-4}
                                                                                           & \textbf{\ac{mAP}}                                                 & \textbf{\ac{MR}}                                                & \textbf{\ac{P@1}}                                                 \\ \hline
    \multicolumn{1}{|l|}{$\mathbf{L_D}$}                                                   & \begin{tabular}[c]{@{}r@{}}81.65\\ ± 0.74\%\end{tabular}          & \begin{tabular}[c]{@{}r@{}}82.57\\ ± 4.74\end{tabular}          & \begin{tabular}[c]{@{}r@{}}80.23\\ ± 0.54\%\end{tabular}          \\ \hline
    \multicolumn{1}{|l|}{$\mathbf{WER}$}                                                   & \begin{tabular}[c]{@{}r@{}}82.91\\ ± 1.27\%\end{tabular}          & \begin{tabular}[c]{@{}r@{}}75.01\\ ± 8.43\end{tabular}          & \begin{tabular}[c]{@{}r@{}}81.52\\ ± 0.97\%\end{tabular}          \\ \hline
    \multicolumn{1}{|l|}{$\mathbf{BOW}$}                                                   & \begin{tabular}[c]{@{}r@{}}85.74\\ ± 0.67\%\end{tabular}          & \begin{tabular}[c]{@{}r@{}}46.29\\ ± 1.80\end{tabular}          & \textbf{\begin{tabular}[c]{@{}r@{}}83.65\\ ± 0.71\%\end{tabular}} \\ \hline
    \multicolumn{1}{|l|}{\textbf{\begin{tabular}[c]{@{}l@{}}Triplet\\ Model\end{tabular}}} & \textbf{\begin{tabular}[c]{@{}r@{}}87.17\\ ± 1.50\%\end{tabular}} & \textbf{\begin{tabular}[c]{@{}r@{}}18.51\\ ± 2.35\end{tabular}} & \begin{tabular}[c]{@{}r@{}}83.57\\ ± 1.80\%\end{tabular}          \\ \hline
  \end{tabular}
  \caption{Average results on the 5-fold cross-validation}
  \label{tab:results}
\end{table}

\section{Limitations and Future Work}

While our approach shows promise, it is not without limitations. Firstly, direct comparisons with other \ac{CSD} approaches that focus on audio features pose challenges, given our method's exclusive operation on our introduced dataset. A meaningful comparison would necessitate additional steps, including downloading and extracting audio files from YouTube, conducting speech-to-text transcriptions, and comparing lyrics. Importantly, a comparison of our approaches to methods that use datasets like \ac{SHS} \cite{SecondHandSongsDatasetMillion} requires evaluating how well current speech-to-text algorithms handle singing voice, potentially involving training models like Whisper \cite{radford2022robust} for this task. This leads to the second limitation, lyrics extracted from the audio may differ from the annotated lyrics, introducing potential errors, or missing annotation. As already mentioned it is necessary to evaluate our approaches on extract lyrics. Lastly, the context window of our model, while suitable for average-sized lyrics, may prove insufficient for larger lyrics, potentially leading to information loss and impacting cover song detection accuracy.

In future work, addressing these limitations is crucial. Expanding the dataset size could enhance the model's performance. Furthermore, we intend to extend training to the entire model, encompassing not only the fully connected and embedding layers. Experimenting with direct lyric extraction from audio, especially from datasets like \ac{SHS}, would facilitate meaningful comparisons with other \ac{CSD} methods. To tackle the context window limitation, research efforts could focus on methods for handling larger lyrics, such as expanding the context window or exploring alternative solutions like splitting lyrics into smaller segments for more effective processing. It might also be possible to partly merge our dataset with other datasets like \ac{SHS} to increase the size of the dataset.

\section{Responsible NLP}

In light of the current state of our methodology, it has been conclusively ascertained that the implementation of a large-scale \ac{CSD} system in a professional environment is currently unfeasible. The metrics provided herein illustrate specific circumstances wherein our methodology fails to produce accurate results, thus introducing a potential risk of incorrect matches. Such inaccuracies could potentially disrupt the fair distribution of royalties. Nevertheless, there are potential solutions that could mitigate this risk. For instance, imposing a distance limit and only considering automated matches within close proximity could be a viable strategy.
We acknowledge that transformer-based models inherently possess biases that may facilitate more effective matching of certain words or phrases over others. Although there is no concrete evidence at present to suggest bias in our model, we unequivocally pledge to comprehensively investigate this issue in our future research endeavors.
With respect to the reproducibility of our approach's results, we do not have concerns. Given that our model does not generate text and as such does not possess a hyperparameter for output generation, our results are reproducible. For the same given lyrics, the same embedding vector will be produced, thereby yielding the identical original lyrics.
In the development of our model, we utilized GitHub Copilot. It is important to note that this tool did not exert any influence on our research, and we carefully reviewed the generated code. Additionally, we employed GPT-4 to check for spelling and grammatical inaccuracies present in our manuscript and correct them.

\bibliography{custom}

\end{document}